\documentclass[final,5p,times,twocolumn]{elsarticle}
\usepackage{amsmath,amssymb,amsfonts}
\usepackage{algorithmic}
\usepackage{graphicx}
\usepackage{hyperref}
\usepackage{textcomp}
\usepackage{xcolor}
\usepackage{booktabs, tabularx, array}
\usepackage[numbers]{natbib}
\usepackage{chngcntr}

\journal{Computers in Biology and Medicine}

\begin{document}

\begin{frontmatter}

\title{AMRG: Extend Vision Language Models for Automatic Mammography Report Generation}

\author[1]{Nak-Jun Sung} 
\author[1]{Donghyun Lee}
% \author[1,3]{Junsang Bae}
\author[2]{Bo Hwa Choi}
\author[1]{Chae Jung Park\corref{cor1}}

% \cortext[cor1]{Corresponding author}
% \ead{cjp@ncc.re.kr}

\cortext[cor1]{Corresponding author. Research Institute, National Cancer Center Korea, Goyang-si, Republic of Korea.\\
E-mail address: \texttt{cjp@ncc.re.kr} (Chae Jung Park).}

%% Author affiliation
\affiliation[1]{organization={Research Institute, National Cancer Center Korea},%Department and Organization
            addressline={323, Ilsan-ro, Ilsandong-gu}, 
            city={Goyang-si},
            postcode={10408}, 
            state={Gyeonggi-do},
            country={Republic of Korea}}

\affiliation[2]{organization={Department of Radiology, National Cancer Center Korea},%Department and Organization
            addressline={323, Ilsan-ro, Ilsandong-gu}, 
            city={Goyang-si},
            postcode={10408}, 
            state={Gyeonggi-do},
            country={Republic of Korea}}

% \affiliation[3]{organization={Deptartment of Computer Software Engineering, Soonchunhyang University},
%             addressline={22, Soonchunhyang-ro, Sinchang-myeon}, 
%             city={Asan-si},
%             postcode={31538}, 
%             state={Chungcheongnam-do},
%             country={Republic of Korea}}

\begin{abstract}
Mammography report generation is a critical yet underexplored task in medical AI, characterized by challenges such as multiview image reasoning, high-resolution visual cues, and unstructured radiologic language. In this work, we introduce \textbf{AMRG} (Automatic Mammography Report Generation), the first end-to-end framework for generating narrative mammography reports using large vision-language models (VLMs). Building upon MedGemma-4B-it—a domain-specialized, instruction-tuned VLM—we employ a parameter-efficient fine-tuning (PEFT) strategy via Low-Rank Adaptation (LoRA), enabling lightweight adaptation with minimal computational overhead. We train and evaluate AMRG on DMID, a publicly available dataset of paired high-resolution mammograms and diagnostic reports. This work establishes the first reproducible benchmark for mammography report generation, addressing a longstanding gap in multimodal clinical AI. We systematically explore LoRA hyperparameter configurations and conduct comparative experiments across multiple VLM backbones, including both domain-specific and general-purpose models under a unified tuning protocol. Our framework demonstrates strong performance across both language generation and clinical metrics, achieving a ROUGE-L score of 0.5691, METEOR of 0.6152, CIDEr of 0.5818, and BI-RADS accuracy of 0.5582. Qualitative analysis further highlights improved diagnostic consistency and reduced hallucinations. AMRG offers a scalable and adaptable foundation for radiology report generation and paves the way for future research in multimodal medical AI.

\end{abstract}

%% Keywords

\end{frontmatter}

\section{Introduction}
%radiology report의 역할 및 비정형 텍스트 자동 생성의 어려움
Generating radiology reports has significant challenges, particularly in the aspect of non-structured text generation. The radiology report encapsulates the core findings of medical image interpretation and serves as a critical communication channel between radiologists and clinicians~\cite{he2025radiology}. It functions as a natural language-based summary that extends beyond mere technical descriptions, exerting a substantial influence on clinical decision-making, from diagnostic confirmation to treatment planning and longitudinal follow-up. Accordingly, the accuracy, clarity, and timeliness of radiology reports are directly associated with patient safety and improved clinical outcomes.

%리포트 작성의 어려움 & 의료 영상 데이터의 증가로 인한 판독 부담 가중
Currently, most radiology reports are manually generated by radiologists following visual analysis of medical images—a process that is both time-consuming and cognitively demanding. In particular, the exponential growth of medical imaging data—driven by the widespread adoption of high-resolution modalities, increased health screening programs, and an aging population—has intensified the interpretative demand on specialists. 

%대규모 판독 수요의 증가와 구조적 일관성 부족으로 인한 mammography report generation 자동화의 필요성
Mammography is a representative image modality where the interpretative demanding is particularly pronounced~\cite{birads}. It serves as a key modality for early breast cancer detection and constitutes a standard procedure in the initial stage of screening programs worldwide. In South Korea, mammography is a biennial mandatory screening for women aged over 40, implemented under the National Cancer Screening Program, with annual examinees numbering in the millions~\cite{koreanbreastcancer2024}. However, the large-scale analysis workload driven by wide early-cancer screening program continues to exceed the capacity of available radiology specialists, leading to clinical challenges such as delayed reporting, missed findings, and diagnostic errors. Accordingly, automated medical image analysis and AI-based report generation are recognized as essential advancements for building a sustainable clinical infrastructure, and standardizing and automating mammography diagnostics reports, which often lack structural consistency, is a promising use case that can simultaneously improve interpretation efficiency and accuracy.

%VLM의 발전으로 인한 medical image-text 정합 가능성 & report generation 특징에 따른 어려움
The recent rapid advancement of Vision-Language Models (VLMs) has enabled sophisticated learning of semantic mappings between medical images and natural language, thereby facilitating active research on end-to-end generation of radiology reports directly from imaging data~\cite{chen2020generating, jin2024promptmrg}. Unlike conventional tasks such as image captioning or visual question answering, medical report generation is inherently more complex and domain-specific, as it requires the production of highly detailed and clinically accurate descriptions. In particular, medical report generation is a high-stakes task, where the choice of a single word can critically affect the clinical interpretation and the overall reliability of the report. For example, the distinction between the terms "normal" and "abnormal" can fundamentally alter the clinical implications, potentially leading to misdiagnosis or inappropriate treatment decisions. 

%의료영상의 구조적 특성(멀티모달, 멀티뷰) & 비정형 리포트 형태로 인한 일관성 확보 어려움
This level of sensitivity distinguishes medical report generation from natural language generation in domains such as general-purpose applications. The structural characteristics of medical images further compound this challenge. Most medical images exhibit low-dimensional, grayscale visual information and are often acquired through multiple sequences (e.g., T1, T2, contrast-enhanced in MRI) or multiple views (e.g., craniocaudal (CC) and mediolateral oblique (MLO) view in mammography). These properties necessitate advanced multimodal and multiview fusion strategies, rather than simple single-image encoding approaches. Moreover, the reports themselves are typically written in unstructured natural language, lacking standardized formatting~\cite{smit2020chexbert}. The choice of terminology and descriptive style can vary significantly depending on the expertise, writing habits, and preferences of the reporting radiologist. For the same finding, terms like "mass", "nodule", and "lesion" are often used interchangeably, introducing inconsistencies that hinder both model training and evaluation. 
%정밀한 병변 인식 및 글로벌 컨텍스트 이해를 위한 모달리티 정렬의 필요성
Additionally, accurate medical report generation requires the integration of both fine-grained local cues—to identify and describe specific lesion findings—and global contextual understanding of the image. To achieve this, models must be equipped with fine-grained visual comprehension capabilities and precise local vision-language alignment mechanisms, enabling them to correctly detect region-specific abnormalities and generate semantically aligned textual descriptions.

\begin{figure}[h]
	\centering	
	\includegraphics[width=\linewidth]{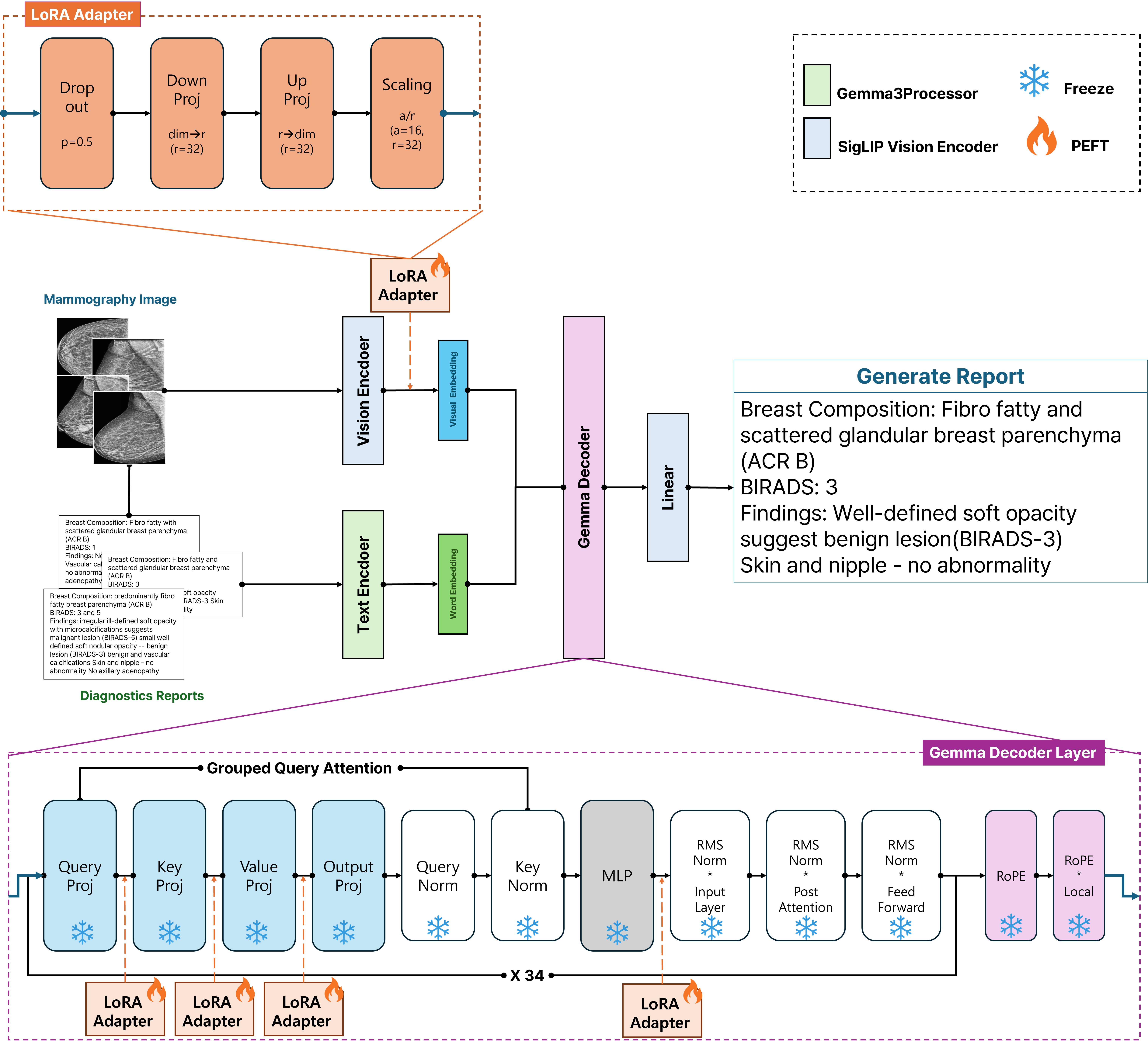}
	\caption{Overview of our proposed Automatic Mammography Report Generation (AMRG) framework. The system leverages the MedGemma-4B vision-language model as a domain-specialized backbone and applies PEFT via LoRA adapters.}	
	\label{fig:Overview}
\end{figure}

%메디컬 도메인에 특화된 VLM(MedGemma-4B-it)에 PEFT 전략을 적용해 mammography domain으로의 확장 진행
Building on the capabilities of MedGemma-4B-it~\cite{sellergren2025medgemma}, a recently released instruction-tuned VLM specialized for the medical domain, we propose an end-to-end framework for \textbf{Automatic Mammography Report Generation (AMRG)}. MedGemma-4B-it combines a SigLIP-based vision encoder with a clinical language model, pre-trained across diverse medical image-text pairs from radiology, dermatology, pathology, and ophthalmology. Leveraging this domain-aware foundation, we extend MedGemma for the mammography domain by introducing a parameter-efficient fine-tuning (PEFT) strategy tailored to the structure and semantics of radiology reports.

%LoRA 적용을 통해 mammo report generation 프레임워크 AMRG를 구축 & NLG, Clinical metric 기반 성능 평가 진행
Specifically, we integrate Low-Rank Adaptation (LoRA)~\cite{hu2022lora} adapters into the linear projection layers of MedGemma’s transformer blocks, enabling efficient adaptation to downstream mammography tasks with minimal computational cost. This design significantly reduces the number of trainable parameters while preserving the general visual-linguistic reasoning capabilities of the backbone. Our AMRG pipeline systematically explores the effects of key LoRA hyperparameters (rank \( r \), scaling factor \( \alpha \), and dropout) on both linguistic quality and clinical accuracy. We evaluate performance using standard natural language generation metrics (e.g., BLEU~\cite{papineni2002bleu}, ROUGE~\cite{lin2004rouge}, METEOR~\cite{banerjee2005meteor}, CIDEr~\cite{vedantam2015cider}) and mammography-specific metrics such as accuracy of Breast Imaging Reporting and Data System (BI-RADS) code and breast density category. This framework not only benchmarks MedGemma’s potential in mammography but also establishes a reusable and efficient protocol for adapting large medical VLMs to other specialized imaging domains.

%연구의 의의(최초의 mammo report generation 벤치마크 제공). 재현 가능한 연구임을 소개
In this study, we introduce a benchmark for automatic radiology report generation in mammography, leveraging the publicly available DMID dataset~\cite{oza2024digital} consisting of paired diagnostic images and narrative reports. Our work establishes a foundation for future multimodal research in this clinically critical yet underrepresented modality. To this end, we design a comprehensive evaluation framework that compares multiple vision-language backbones—including domain-specific (MedGemma) and general-purpose (Qwen2.5-VL, Phi-3.5-VL) models, as well as modular architectures (CLIP and MedCLIP with GPT2 decoders)—under consistent PEFT setups. We further investigate the effects of prompt design and fine-tuning configurations on both linguistic quality and clinical correctness, offering holistic insights into model behavior. By standardizing inputs, outputs, and evaluation criteria, our benchmark facilitates reproducible research and establishes a clear baseline for future improvements in mammography-specific report generation.
\section{Related Works}
\subsection{Vision-Language Models in the Medical Domain}
%멀티모달 특성을 가진 의료 도메인과 VLM의 적합성 & 대표적인 task인 report generation 소개
The medical domain, with its inherently multimodal nature of clinical decision-making, encompasses several application areas well suited for vision–language models. Among these, diagnostic report generation is particularly aligned with the strengths of VLMs, as it inherently involves interpreting medical images and composing corresponding narrative reports.

%초기 report generation 연구 방법 소개
Initial efforts adapted general-purpose architectures by pairing standard vision backbones (e.g., ResNet, ViT) with pretrained language models (e.g., BERT, GPT), then fine-tuning them on small-scale medical corpora. Representative examples include BioViL~\cite{bannur2023learning}, MedCLIP~\cite{wang2022medclip}, and GLoRIA~\cite{huang2021gloria}, which applied contrastive learning on paired image-report datasets such as MIMIC-CXR~\cite{johnson2019mimic}. These approaches demonstrated improvements in classification and retrieval, yet lacked generative capabilities necessary for free-text report synthesis.

%최근 report generation 연구 방법 소개
To address this, recent works have transitioned toward generative VLMs, many of which leverage instruction tuning. For example, LLaVA-Med~\cite{li2023llava} extends the LLaVA~\cite{liu2023visual} framework by aligning general-purpose VLMs to biomedical tasks via continued pretraining and multimodal instruction tuning. It supports tasks such as VQA, image captioning, and limited-form report generation, though typically within simplified domains like chest X-ray (CXR) interpretation. While instruction-following capabilities have expanded the model’s generalizability, full adaptation to complex and underrepresented imaging modalities remains limited.
%멀티모달 파운데이션 모델 RadFM에 대한 자세한 설명 & 해당 모델이 가진 mammography 분야의 한계점 소개
RadFM~\cite{wu2023towards} represents a multimodal foundation model designed for radiology-specific tasks. Trained through a staged process—combining masked image modeling, vision-language contrastive learning, and instruction tuning—RadFM supports a variety of downstream tasks, including tagging, VQA, and summarization. Importantly, it incorporates diverse imaging modalities, including mammography via datasets such as VinDr-Mammo. However, its use of mammographic data is restricted to structured tasks like lesion tagging, as VinDr-Mammo lacks diagnostic reports. Consequently, RadFM does not address the challenge of generating full free-text mammography reports.

\subsection{Automatic Radiology Report Generation}
%의료 영상 기반의 캡셔닝을 위한 기존 연구 방향 & 최근 연구 트렌드 전환 과정
Research on report generation for medical images began as an extension of the existing natural image captioning method using a combination of encoders and decoders~\cite{hossain2019comprehensive, you2016image}. 
However, captioning medical images is more difficult than captioning natural images. To solve this limitation, various studies are being conducted, such as strengthening the encoder model~\cite{chen2020generating, yuan2019automatic}, changing the decoder using LLM~\cite{wang2023r2gengpt, pellegrini2023radialog, he2025radiology},and compact models trained via knowledge distillation~\cite{khan2026radiology}. Through these techniques, automatic radiology report generation tasks for various medical modalities have been rapidly developed.

%Chest X-Ray에 대한 최근 연구 조사 진행
\paragraph{Chest X-ray} CXR interpretation has advanced significantly, driven by the availability of large-scale datasets with structured and narrative annotations—such as MIMIC-CXR~\cite{johnson2019mimic} and CheXpert~\cite{irvin2019chexpert}. Sîrbu et al.~\cite{sîrbu2025gitcxrendtoendtransformerchest} propose GIT‐CXR, an end‐to‐end Transformer augmented with curriculum learning, setting new state-of-the-art performance on METEOR and clinical accuracy metrics (F1‐micro/macro/example‐averaged) on MIMIC‐CXR. Singh et al. ~\cite{singh2025chestx} propose a ChestX‐Transcribe that combines Swin-Transformer for high‐resolution visual encoding with DistilGPT for clinical text generation, outperforming prior models on BLEU, ROUGE, and METEOR in the IU chest X‐ray dataset. Liu et al.~\cite{liu2025enhanced} introduce MLRG, leveraging multi‐view longitudinal contrastive pretraining and tokenized absence encoding, improving BLEU‐4 (+2.3\%), F1 (+5.5\%), and RadGraph F1 (+2.7\%) over SOTA on MIMIC‑CXR, MIMIC‑ABN and two‑view CXR benchmarks.

%MRI, Pathology에 대한 최근 연구 조사 진행
\paragraph{MRI and Pathology} In addition to CXR, report generation research has been extended to other medical imaging modalities, including pathology and magnetic resonance imaging (MRI). This reflects a growing interest in developing modality-specific generative frameworks tailored to the distinct visual and linguistic characteristics of each domain.
BiGen~\cite{zhang2025historical} proposes a Historical Report Guided Bi-modal Concurrent Learning Framework that enriches Whole Slide Image encodings with retrieved semantic knowledge, achieving a 7.4\% relative improvement in NLP metrics and a 19.1\% boost in HER-2 classification on the PathText (BRCA) dataset. AutoRG-Brain~\cite{lei2024autorg} introduces the first brain MRI report generator grounded in pixel‑level visual cues and trained on the new RadGenome‑Brain MRI dataset. Their study extracts grounded masks (local masks) using a high-performance segmentation model and uses them as input to perform report generation. By utilizing the high-performance segmentation results, leading to improved performance in global report generation.

\subsection{Mammography Report Generation}
%mammography 분야의 VLM 적용 연구가 적은 이유 / 현재 public 데이터셋의 한계점
Mammography remains significantly underexplored within the VLM literature, largely due to the scarcity of publicly available datasets that contain both high-resolution screening images and corresponding narrative reports. While datasets such as DDSM~\cite{ddsm} and VinDr-Mammo~\cite{VinDr} offer diagnostic labels (e.g., BI-RADS code and breast density category), lesion masks, and metadata, they lack the radiologist-written textual reports required to train generative models. Consequently, most prior work in this modality has focused on classification or detection tasks, with limited attention given to language generation.
%기존 연구 bertmammo에 대한 소개 / private 데이터셋 사용으로 인한 재현성 제한에 대한 설명
Among the few existing studies, Yalunin et al.~\cite{yalunin2021bertmammo} proposed one of the earliest models for automated report generation from multi-view mammograms. Their architecture combines an EfficientNet-based encoder with a Transformer-based decoder, leveraging attention mechanisms to localize salient image regions and generate narrative reports. Clinical evaluation by a certified radiologist demonstrated the potential of their approach. However, this work relied on a proprietary dataset curated from the Russian national breast cancer screening program, limiting reproducibility and hindering fair benchmarking by the broader community.
%DMID 데이터셋을 이용한 한계점 개선
To address this gap, we leverage the recently released Digital Mammography Dataset for Breast Cancer Diagnosis Research (DMID)~\cite{oza2024digital}, which includes high-resolution mammograms (in DICOM and TIFF formats) and radiologist-authored narrative reports. The dataset also provides region-of-interest (ROI) masks and structured metadata, enabling comprehensive multimodal learning for clinically grounded report generation.
%기존 연구와 다른 본 연구의 접근 방향 제시
In contrast to previous studies that either target structured prediction tasks (e.g., classification or detection) or rely on private datasets, our work uniquely addresses the underexplored challenge of free-text mammography report generation in a fully multimodal setting. Leveraging MedGemma, a medical-domain VLM, we propose a PEFT strategy based on LoRA, applied to each linear projection layer in the model. This allows for efficient adaptation to the downstream task of narrative report generation with minimal computational burden. Beyond adopting LoRA, we systematically explore its hyperparameter configurations—such as rank and scaling factor—and evaluate their effect on both linguistic and clinical quality using a comprehensive suite of evaluation metrics. To the best of our knowledge, this study represents the first application of an instruction-tuned, domain-specialized VLM to the task of mammography report generation on a publicly available paired image–text dataset, thus establishing a reproducible benchmark for future research in this domain.
\section{Method}
\subsection{Data Curation and Preprocessing}
\begin{figure}[ht]
	\centering	
	\includegraphics[width=\linewidth]{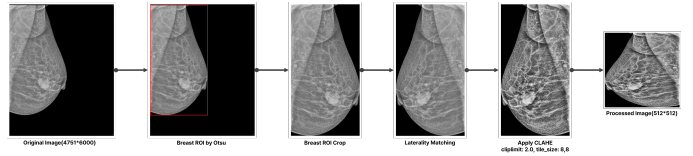}
	\caption{Stages of the Mammography Image Preprocessing Pipeline.}	
	\label{fig:Preporcessing_Image}
\end{figure}

%DMID 데이터셋에 대한 정리(데이터 갯수, 분할, 특성 소개 등)
We leverage the DMID dataset, which contains 510 annotated mammography cases comprising high-resolution images paired with radiologist-generated diagnostic reports. We follow a three-way split, with 407 cases in the training set, 51 in validation, and 52 in the test set. The distribution of BI-RADS categories is notably imbalanced, reflecting the real-world prevalence of benign findings in screening populations. Most cases are labeled as BI-RADS 1 (negative) or BI-RADS 3 (probably benign), while high-suspicion categories such as BI-RADS 4b, 4c, and 5 appear less frequently, though in meaningful proportions within the training set. The validation and test splits contain a balanced mix of benign and suspicious cases, enabling robust evaluation across a range of diagnostic scenarios. A small number of ambiguous or mixed labels (e.g., "3 and 5") are also present. This class imbalance poses challenges for both classification and report generation tasks, as models may overfit to dominant categories or fail to capture clinically significant but underrepresented patterns. The full distribution of BI-RADS codes across splits as shown in Table~\ref{tab:birads_split} in Appendix.
%이미지 전처리를 위한 processing 단계 소개
To standardize and enhance the mammography images prior to training, we implement a multi-stage preprocessing pipeline designed to improve visual quality and anatomical alignment while reducing irrelevant background regions. First, to isolate the breast region from the high-resolution mammogram, we apply Otsu’s thresholding algorithm~\cite{otsu1979threshold} to the grayscale version of the image. This adaptive method selects an optimal intensity threshold that separates foreground (breast tissue) from background. A tight bounding box is then fitted around the resulting binary mask to crop the region of interest (ROI), effectively removing large blank margins. This cropped region is subsequently resized to a fixed resolution of \(512 \times 512\) pixels to ensure consistency across all samples. Second, to enforce anatomical consistency in left-right orientation, we horizontally flip the image when the breast laterality is labeled as “left,” such that all breasts are oriented to face right. This standardization mitigates directional bias during training and improves generalization across views. Finally, to enhance local contrast and improve lesion visibility, we apply Contrast Limited Adaptive Histogram Equalization (CLAHE) in the LAB color space. Specifically, we use a tile grid size of \(8 \times 8\) and a clip limit of 2.0. This transformation equalizes local brightness within small image tiles while suppressing noise amplification in homogeneous regions. Together, these steps yield a robust and uniform image representation suitable for instruction-tuned report generation. Figure~\ref{fig:Preporcessing_Image} illustrates each stage of the preprocessing pipeline described above. From left to right, the figure shows the original mammogram, the Otsu-thresholded binary mask with ROI cropping, the left-right aligned breast image, and the final contrast-enhanced image after CLAHE. This visualization highlights the impact of each step on improving anatomical clarity, reducing background noise, and standardizing the input space for training.

\subsection{Low-Rank Adaptation of Vision-Language Model}
%MedGemma-4B-it에 LoRA를 전면 적용해 경량화 진행 & 효율적 mammo 보고서 생성 파인튜닝 / 하이퍼파라미터 탐색 방법 소개
To adapt the MedGemma-4B-it to the mammography report generation task, we employ parameter-efficient fine-tuning using LoRA~\cite{hu2022lora}. LoRA introduces a trainable low-rank update to linear transformations, allowing for effective adaptation without modifying the full set of pretrained weights.

Let \( W \in \mathbb{R}^{d \times k} \) denote a weight matrix in the original model. Instead of updating \( W \) directly, LoRA learns an additive perturbation \( \Delta W \) expressed as a product of two low-rank matrices:

\begin{equation}
\Delta W = A B, \quad A \in \mathbb{R}^{d \times r}, \quad B \in \mathbb{R}^{r \times k}, \quad r \ll \min(d, k)
\end{equation}

where \( r \) is the rank of the adaptation. The adapted weight matrix is then given by:

\begin{equation}
W' = W + \alpha \cdot \Delta W = W + \alpha A B
\end{equation}

with \( \alpha \in \mathbb{R} \) being a scaling factor that modulates the impact of the low-rank update.

In our implementation, LoRA modules are inserted into all linear layers across the MedGemma architecture, spanning both the encoder and decoder. Specifically, they are applied to the attention projection layers (e.g., query, key, value, and output), the feed-forward network layers (e.g., first linear, second linear, and gating projection), and the gated MLP components (e.g., up, down, and output projections). This comprehensive injection strategy enables flexible yet efficient adaptation throughout the model while keeping all pretrained weights frozen.

To explore the impact of LoRA capacity, we conduct a grid search over rank values \( r \in \{16, 32, 64\} \) and scaling factors \( \alpha \in \{8, 16\} \). A dropout of 0.05 is applied to the LoRA modules to improve generalization. During fine-tuning, only the LoRA parameters, embedding layer (\texttt{embed\_tokens}), and output head (\texttt{lm\_head}) are updated; all other parameters remain frozen. An overview of the architecture and adaptation strategy is illustrated in Figure~\ref{fig:Overview}.

\subsection{Loss for VLM}
%2가지 방식(LoRA 어댑터 추가 / Module Custom Decoder)에 따른 Loss 설명 & 수식 유도

To train the VLMs for radiology report generation, we apply a conditional language modeling loss that accounts for both textual and visual modalities. Depending on the architecture of the model—monolithic multimodal transformers (e.g., MedGemma, Qwen-VL, Phi) versus modular CLIP+decoder pipelines—we adopt different loss formulations tailored to their decoding mechanisms.

\paragraph{Casual LM Loss for Instruction-Tuned Multimodal LLMs}
For unified vision-language backbones such as MedGemma-4B-it, Qwen2.5-VL-7B, and Phi-3.5-VL, we adopt a causal language modeling (CLM) loss conditioned on visual input and task-specific instructions. Given an input image \( I \), the visual encoder extracts a feature embedding \( \mathbf{v} = \texttt{ImageEncoder}(I) \). This embedding is combined with the tokenized instruction prompt \( \mathbf{x}^{\text{inst}} = \{x_1, \dots, x_M\} \) to form the model input sequence, either via prepending (e.g., MedGemma) or token interleaving (e.g., Qwen2.5-VL). The language decoder then autoregressively generates the target report sequence \( y = \{y_1, \dots, y_T\} \).

The conditional probability of the report sequence, given the visual context and instructions, is factorized as:
\begin{equation}
P(y \mid \mathbf{v}, \mathbf{x}^{\text{inst}}; \theta) = \prod_{t=1}^{T} P(y_t \mid y_{<t}, \mathbf{v}, \mathbf{x}^{\text{inst}}; \theta),
\end{equation}
where \( \theta \) denotes the model parameters. Accordingly, the CLM loss is defined as the average negative log-likelihood over the output sequence:
\begin{equation}
\mathcal{L}_{\text{CLM}} = - \frac{1}{T} \sum_{t=1}^{T} \log P(y_t \mid y_{<t}, \mathbf{v}, \mathbf{x}^{\text{inst}}; \theta).
\end{equation}

During training, we apply teacher forcing, where the decoder is conditioned on the ground-truth prefix \( y_{<t} \) at each time step. This objective encourages the model to generate semantically consistent and visually grounded medical reports by leveraging both the multi-view imaging context and instruction-driven prompts.

\paragraph{Cross-Attentive Decoder Loss for CLIP-based Models}
In contrast to unified VLMs, CLIP-based architectures such as CLIP+GPT2 and MedCLIP+GPT2 follow a modular design in which a pretrained image encoder produces visual features \( \mathbf{v}_i \in \mathbb{R}^{L \times d_v} \) for the \( i \)-th image. These features are injected into a GPT2-style decoder via cross-attention modules at each transformer block.

Let \( y_i = \{y_{i,1}, \dots, y_{i,T}\} \) denote the tokenized report sequence for the \( i \)-th sample. At each decoding step \( t \), the hidden representation \( h_t \) is obtained via masked self-attention followed by cross-attention with the visual context:

\begin{equation}
h_t = \mathrm{CrossAttn}(\mathrm{SelfAttn}(y_{i,<t}), \mathbf{v}_i)
\end{equation}

The cross-attention operation computes the attended output using projected query-key-value matrices as:

\begin{equation}
\mathrm{CrossAttn}(h_t, \mathbf{v}_i) = \mathrm{Attention}(Q, K_v, V_v)
\end{equation}
\begin{equation}
Q = W^Q h_t,\quad K_v = W^K \mathbf{v}_i,\quad V_v = W^V \mathbf{v}_i
\end{equation}

\begin{equation}
\mathrm{Attention}(Q, K_v, V_v) = \mathrm{softmax}\left( \frac{Q K_v^\top}{\sqrt{d_k}} \right) V_v
\end{equation}

The final token probability is computed by projecting the attended hidden state:

\begin{equation}
P(y_{i,t} \mid y_{i,<t}, \mathbf{v}_i) = \mathrm{softmax}(W_o h_t + b)[y_{i,t}]
\end{equation}

We define the overall CLM loss for CLIP-based models as $\mathcal{L}_{\text{GPT2}}$, which is given by:

\begin{equation}
\mathcal{L}_{\text{GPT2}} = -\frac{1}{N T} \sum_{i=1}^{N} \sum_{t=1}^{T} \log P(y_{i,t} \mid y_{i,<t}, \mathbf{v}_i)
\end{equation}

As with instruction-tuned models, we apply teacher forcing during training. The decoder learns to align visual embeddings with textual outputs by attending to image features at every decoding layer, facilitating effective multimodal grounding.

\section{Experiments}
%제안한 AMRG의 성능을 검증하기 위해 LoRA ablation, VLM 백본 비교 등 실험을 설계
In this section, we present a comprehensive set of experiments designed to evaluate the effectiveness of our proposed MedGemma-based report generation framework for mammography. Our study aims to establish a strong baseline for this underexplored task by systematically analyzing key components that influence performance. We organize the experiments into three primary categories: 

%LoRA Ablation의 configuration 소개
\paragraph{LoRA Configuration Ablation} We investigate how the choice of LoRA parameters—namely the rank ($r \in \{16, 32, 64\}$) and scaling factor ($\alpha \in \{8, 16\}$)—affects report generation quality. This allows us to characterize the trade-off between model expressiveness and parameter efficiency under a parameter-efficient fine-tuning scheme. 
%VLM 백본 ablation의 configuration 소개
\paragraph{VLM Backbone Ablation}
To examine the effect of backbone architecture on mammography report generation, we compare four VLMs under a unified fine-tuning setting: MedGemma-4B (proposed), Qwen2.5-VL-7B~\cite{bai2025qwen25vltechnicalreport}, Phi-3.5-Vision~\cite{abdin2024phi3technicalreporthighly}, CLIP~\cite{clip} + GPT2Decoder, and MedCLIP~\cite{wang2022medclip} + GPT2Decoder. All models are fine-tuned using LoRA adapters with identical hyperparameters (\( r=32 \), \( \alpha=16 \), \( \tau=0.1 \)) and trained on the DMID dataset. This comparison allows us to evaluate the role of domain specialization, model scale, and modularity in radiology-oriented text generation. A detailed analysis of each model’s performance—across both standard NLP metrics and clinically grounded label accuracies—is presented in subsequent sections.

\subsection{LoRA Configuration Ablation}
%LoRA rank·scaling factor 변경에 따른 성능 비교. r(16, 32, 64), α(8, 16)
To investigate the sensitivity of our model to different parameter-efficient fine-tuning setups, we conduct a series of ablation experiments on the LoRA configuration. In particular, we vary two key hyperparameters: the rank $r \in \{16, 32, 64\}$ of the low-rank decomposition and the scaling factor $\alpha \in \{8, 16\}$ applied to the residual adapter. These parameters govern the representational capacity of the LoRA modules and their contribution to the final output. By systematically sweeping across the configuration space, we aim to understand the trade-off between adaptation strength and overfitting risk in the context of mammography report generation. Each configuration is trained on the DMID dataset with identical training settings: 20 epochs, a batch size of 4, AdamW optimizer with a learning rate of 1e-4, and gradient accumulation steps of 8. In addition to standard NLP metrics, we evaluate BI-RADS category and breast density prediction as multi-class classification tasks, where accuracy is computed as the proportion of exact matches between predicted and ground-truth labels.

\begin{table}[ht]
\centering
\caption{Evaluation results of MedGemma-4B fine-tuned with various LoRA configurations on the DMID dataset. The baseline performance using original MedGemma-4B~\cite{sellergren2025medgemma} without finetuning is compared against six finetuning options with varying LoRA ranks (\( r \in \{16, 32, 64\} \)) and scaling factors (\( \alpha \in \{8, 16\} \)). All generations are performed with temperature \( \tau = 0.1 \). Best results per row are underlined and bolded.}
\label{tab:ablation_lora_metrics}
\resizebox{\linewidth}{!}{
\begin{tabular}{l|c|cccccc}
\toprule
\textbf{Metric} 
& \textbf{Baseline} 
& \textbf{$r=16, \alpha=8$} 
& \textbf{$r=32, \alpha=8$} 
& \textbf{$r=64, \alpha=8$} 
& \textbf{$r=16, \alpha=16$} 
& \textbf{$r=32, \alpha=16$} 
& \textbf{$r=64, \alpha=16$} \\
\midrule
BLEU-1 & 0.0025 & 0.1870 & 0.2550 & 0.2449 & 0.2223 & \underline{\textbf{0.3075}} & 0.2694 \\
ROUGE-1 & 0.0684 & 0.4657 & 0.5305 & 0.5166 & 0.5119 & \underline{\textbf{0.5750}} & 0.5280 \\
ROUGE-2 & 0.0082 & 0.2721 & 0.3449 & 0.3314 & 0.3095 & \underline{\textbf{0.3980}} & 0.3522 \\
ROUGE-L & 0.0613 & 0.4513 & 0.5198 & 0.5032 & 0.4968 & \underline{\textbf{0.5691}} & 0.5188 \\
METEOR & 0.1000 & 0.5193 & 0.5762 & 0.5433 & 0.5541 & \underline{\textbf{0.6152}} & 0.5608 \\
CIDEr & 0.1745 & 0.4827 & 0.5426 & 0.5173 & 0.5180 & \underline{\textbf{0.5818}} & 0.5378 \\
F1 (word-level) & 0.0636 & 0.4537 & 0.5168 & 0.4969 & 0.4978 & \underline{\textbf{0.5610}} & 0.5195 \\
\midrule
Density Accuracy & 0.0000 & 0.4510 & 0.3922 & 0.3137 & \underline{\textbf{0.4902}} & 0.3529 & 0.3725 \\
BI-RADS Accuracy & 0.0000 & 0.4418 & \underline{\textbf{0.5686}} & 0.5294 & 0.3529 & 0.5582 & 0.5490 \\
\bottomrule
\end{tabular}
}
\end{table}

%실험 결과에 대한 간단한 소개
Table~\ref{tab:ablation_lora_metrics} summarizes the results across seven NLP metrics and two clinical classification metrics. We observe that LoRA configurations with moderate rank and scaling factors yield the best overall performance. In particular, the configuration $(r=32, \alpha=16)$ achieves the highest scores across all NLP metrics (e.g., ROUGE-L 0.52, METEOR 0.5194, CIDEr 0.5336) and clinical metrics (BI-RADS accuracy 0.55, density accuracy 0.35), outperforming both the base model and other LoRA variants.

Interestingly, increasing the rank to $r=64$ leads to degraded performance despite higher representational capacity. This suggests that larger LoRA modules may induce overfitting on relatively small datasets like DMID. Conversely, lower-rank settings such as $(r=16, \alpha=16)$ offer competitive results while maintaining lower parameter overhead, making them suitable for deployment scenarios with limited compute.

\subsection{VLM Backbone Ablation}
%VLM 백본 변경에 따른 성능 비교 실험
To assess the impact of backbone architecture on mammography report generation, we compare five VLMs: \textbf{MedGemma-4B} (our proposed model), Qwen2.5-VL-7B, Phi-3.5-4.2B, CLIP+GPT2 Decoder, and MedCLIP+GPT2 Decoder. Building on the findings from our LoRA configuration ablation study (Table~\ref{tab:ablation_lora_metrics}), where the optimal hyperparameters were determined to be (\( r = 32 \), \( \alpha = 16 \), \( \tau = 0.1 \)), all models are fine-tuned under this identical parameter-efficient setup on the DMID dataset.

\begin{table}[ht]
\centering
\caption{Performance comparison of radiology report generation across five VLMs on the DMID dataset. All models are fine-tuned under identical parameter-efficient setups using LoRA adapters with rank \( r = 32 \), scaling factor \( \alpha = 16 \), and temperature \( \tau = 0.1 \). Evaluation includes standard NLP generation metrics (BLEU, ROUGE, METEOR, CIDEr, word-level F1) and clinical classification metrics (BI-RADS accuracy and breast density accuracy). The best value for each metric is underlined and bolded.}
\label{tab:vlm_comparison}
\resizebox{\linewidth}{!}{%
\begin{tabular}{l|c|cccc}
\toprule
\textbf{Evaluation Metric} & \textbf{MedGemma-4B}& \textbf{Qwen2.5-VL-7B}  &\textbf{Phi-3.5-4.2B}&\textbf{CLIP} 
&\textbf{MedCLIP}\\
\midrule
BLEU-1 & 0.3075 & \underline{\textbf{0.3212}}  &0.0880&0.1462 
&0.2202\\
ROUGE-1 & \underline{\textbf{0.5750}} & 0.5685  &0.3673&0.4840 
&0.4983\\
ROUGE-2 & 0.3980 & \underline{\textbf{0.4103}}  &0.1736&0.3181 
&0.3353\\
ROUGE-L & \underline{\textbf{0.5691}} & 0.5634  &0.3559&0.4778 
&0.4891\\
METEOR & \underline{\textbf{0.6152}} & 0.5803  &0.3783&0.4570 
&0.5371\\
CIDEr & \underline{\textbf{0.5818}} & 0.5627  &0.3540&0.3890 
&0.4740\\
F1 (word-level) & \underline{\textbf{0.5610}} & 0.5509  &0.3367&0.4050 
&0.4831\\
\midrule
Density Accuracy & 0.3529 & \underline{\textbf{0.4510}}  &0.2745&0.1176 
&0.1176\\
BI-RADS Accuracy & \underline{\textbf{0.5582}} & 0.4510  &0.1176&0.3333 &0.4902\\
\bottomrule
\end{tabular}
}
\end{table}

%VLM 성능 비교에 대한 간단한 설명
As presented in Table~\ref{tab:vlm_comparison}, \textbf{MedGemma-4B} demonstrates superior performance across six of nine evaluation metrics, notably excelling in ROUGE-1 (0.5750), ROUGE-L (0.5691), METEOR (0.6152), CIDEr (0.5818), word-level F1 score (0.5610), and BI-RADS accuracy (0.5582). These metrics collectively reflect the model's ability to generate reports that are both semantically rich and clinically aligned. While \textbf{Qwen2.5-VL-7B} slightly outperforms MedGemma in BLEU-1 (0.3212 vs. 0.3075) and ROUGE-2 (0.4103 vs. 0.3980), these gains are marginal and confined to surface-level n-gram overlap. In contrast, MedGemma's higher METEOR and CIDEr scores indicate stronger fluency, lexical diversity, and alignment with clinically informative content.

Although Qwen2.5-VL-7B attains the highest breast density classification accuracy (0.4510), it falls short in the more critical BI-RADS prediction task (0.4510 vs. 0.5582), underscoring limitations in clinical reasoning. The discrepancy highlights that general-purpose VLMs, even when scaled up to 7B parameters, may struggle with nuanced diagnostic generation tasks absent domain-specific pretraining.

Performance from the lightweight \textbf{CLIP+GPT2 Decoder} baseline further illustrates this point, with substantial degradation in both language quality (e.g., CIDEr 0.3890) and clinical metrics (BI-RADS accuracy 0.3333). The \textbf{Phi-3.5-4.2B} model similarly underperforms across all axes, with notably low BI-RADS accuracy (0.1176), suggesting that compact generalist VLMs lack the representational grounding necessary for expert-level clinical text synthesis.

\begin{figure*}[t]
	\centering	
	\includegraphics[width=\textwidth]{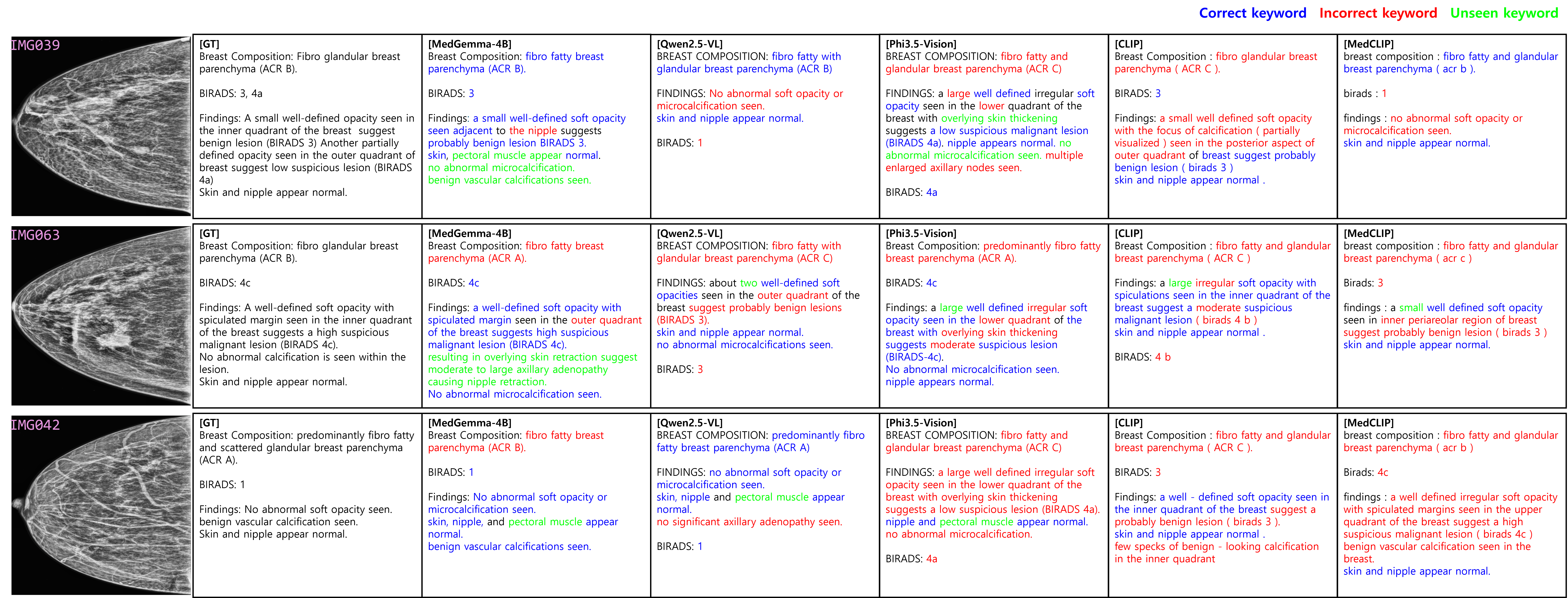}
	\caption{
Qualitative comparison of generated mammography reports across five VLMs: MedGemma-4B, Qwen2.5-VL, Phi-3.5-Vision, CLIP+GPT2 and MedCLIP+GPT2. For each model, generated outputs are aligned with the corresponding ground-truth report. Clinical terms correctly generated are highlighted in \textcolor{blue}{blue}, incorrect terms in \textcolor{red}{red}, and hallucinated or unseen terms (not present in the ground truth) in \textcolor{green}{green}. This visualization illustrates differences in diagnostic accuracy, factual consistency, and content relevance across model architectures.
	}
	\label{fig:experiments}
\end{figure*}

%도메인 특화 모델인 MedGemma의 좋은 성능에 대한 분석
These results collectively demonstrate domain specialization through medical pretraining and instruction tuning, as embodied by MedGemma-4B, is critical for high-fidelity radiology report generation. Despite its smaller scale, MedGemma surpasses the larger Qwen2.5-VL-7B, reinforcing that architectural alignment with clinical priors is more impactful than sheer model size in medical vision-language tasks.

%정성적 평가에 따른 분석 진행
Figure~\ref{fig:experiments} presents representative examples of generated reports from five models: MedGemma-4B, Qwen2.5-VL, Phi-3.5-Vision, CLIP+GPT2, and MedCLIP+GPT2. To facilitate clinical interpretation, we annotate key terms in each generated report using color-coded highlights: \textcolor{blue}{correct terms}, \textcolor{red}{incorrect or misleading terms}, and \textcolor{green}{hallucinated or unseen terms}. Among the compared models, MedGemma-4B demonstrates superior ability to identify and describe salient radiologic findings—such as “spiculated mass” and “architectural distortion”—across multi-view inputs. Its outputs exhibit strong coherence and contextual integration, rarely contradicting the ground-truth interpretation. Although occasional hallucinations of benign anatomical structures are observed, the model maintains high clinical trustworthiness overall. By contrast, Qwen2.5-VL-7B produces syntactically fluent and well-formed radiological sentences but frequently omits or simplifies critical details (e.g., lesion margins, microcalcifications) and intermittently hallucinates unsupported findings (e.g., “calcified lymph node”), highlighting the need for rigorous post-hoc fact verification before clinical deployment. Phi3.5-Vision exhibits frequent misclassification of BI-RADS categories and resorts to generic, non-specialized terminology (e.g., “large irregular soft opacity”), thereby undermining its utility for diagnostic reporting. The CLIP + GPT2 baseline achieves only minimal correct keyword reproduction and is characterized by pervasive inaccuracies and hallucinations, indicating it is unsuitable for preliminary clinical use. Finally, MedCLIP + GPT2 demonstrates modest gains in capturing certain descriptors (e.g., “fibro-fatty parenchyma”) relative to CLIP + GPT2, yet it still suffers from high error rates in BI-RADS prediction and lesion characterization, indicating substantial room for improvement.
\section{Discussion}

\subsection{Observed Challenges in Mammography Report Generation}
%mammography 보고서 생성의 어려움에 대한 디스커션 섹션. 기존 내용은 introduction 내용과 너무 중복되어, 정리하였습니다.
Our experiments with the proposed AMRG framework reveal several inherent challenges in mammography report generation that extend beyond the well-known scarcity of paired image–text datasets. 
First, key clinical labels such as BI-RADS category and breast density are partially subjective, with interpretations varying across radiologists; this subjectivity directly impacts model training stability and leads to substantial variance in generation quality. 
Second, the creation of high-quality, paired mammography datasets is inherently difficult due to the need for expert annotation, privacy concerns, and multi-center data harmonization, making large-scale, diverse corpora rare. 
Third, even when identical BI-RADS labels are provided, narrative reports often contain widely varying lexical and descriptive choices for the same lesion type (e.g., interchangeable use of “mass,” “nodule,” and “lesion”), increasing the complexity of language modeling and evaluation. 
Finally, objective quantification of report quality remains difficult—standard NLP metrics capture surface-level similarity but fail to fully reflect clinical correctness or the nuanced reasoning expected in radiology reporting. 
These factors, confirmed through our ablation and backbone comparison results, highlight that the underrepresentation of mammography report generation in VLM research stems not only from data scarcity but also from intrinsic modality-specific ambiguities and evaluation challenges.

\subsection{Ablation Experiment Results Analysis}
%Ablation Study에 대한 상세한 분석 진행

\paragraph{\textbf{LoRA Configuration}}
%LoRA 어댑터 파라미터 변경에 따른 Ablation에 대한 분석
We evaluate the impact of LoRA adapter hyperparameters on report generation quality by sweeping rank \(r \in \{16,32,64\}\) and scaling factor \(\alpha \in \{8,16\}\) (Table~\ref{tab:ablation_lora_metrics}). All models are fine-tuned for 20 epochs on DMID with identical training settings, and compared against the frozen MedGemma-4B baseline. The configuration \((r=32,\alpha=16)\) yields the strongest overall language performance, achieving BLEU-1 of 0.3075, ROUGE-1 of 0.5750, ROUGE-2 of 0.3980, ROUGE-L of 0.5691, METEOR of 0.6152, CIDEr of 0.5818 and word-level F1 of 0.5610, representing a dramatic improvement over the near-zero baseline. The mid-capacity adapter also produces competitive clinical label accuracy (BI-RADS 0.5582, density 0.3529), confirming its balanced expressiveness. Clinical metrics exhibit slightly different optima: the highest BI-RADS accuracy (0.5686) occurs at \((r=32,\alpha=8)\), while the best density classification (0.4902) is attained at \((r=16,\alpha=16)\). 
%LoRA parameter 변화에 따라 성능의 편차가 발생하는데, 어떤 이유로 발생하는지 분석.
These resurts are strongly influenced by the limited size of DMID. Increasing the rank \(r\) enlarges the number of trainable parameters, which can enhance representational capacity but also raises the risk of overfitting—an effect that becomes pronounced with small datasets. In our experiments, \(r=64\) consistently degraded performance across both NLP and clinical metrics, suggesting that the model began to memorize training-specific patterns rather than generalizing to unseen cases. Conversely, too small a rank (e.g., \(r=16\)) limits capacity but, when paired with an adequate scaling factor (\(\alpha=16\)), can still yield competitive results while avoiding overfitting.
%소규모 데이터셋에서는 중간 수준의 rank, scailing factor를 설정하는 것이 성능에 좋다는 분석
The scaling factor \(\alpha\) controls how strongly the LoRA updates influence the final weights. On a small dataset, a very low \(\alpha\) (e.g., \(\alpha=8\)) can underutilize the limited learning signal available, especially for fine-grained clinical descriptors, leading to underfitting. A moderate \(\alpha\) (e.g., \(\alpha=16\)) better amplifies the adaptation without overwhelming the pretrained backbone, striking a balance between learning new domain-specific patterns and preserving general visual–linguistic reasoning. Overall, our findings indicate that with small datasets like DMID, moderate \(r\) and \(\alpha\) values provide the most stable trade-off between model capacity and the risk of overfitting or underfitting.

\paragraph{\textbf{VLM Models}}
%DMID 실험에서 성능 순서 분석
To assess the role of backbone architecture in mammography report generation, we fine-tuned five vision–language models under an identical LoRA setup (\(r=32,\alpha=16,\tau=0.1\)) on the DMID dataset and report results in Table~\ref{tab:vlm_comparison} and Figure~\ref{fig:experiments}. Quantitatively, the overall performance follows a consistent hierarchy: \textbf{medical-domain specialized VLM} (MedGemma-4B) $>$ \textbf{high-quality general-purpose VLM} (Qwen2.5-VL-7B) $>$ \textbf{custom modular VLMs} (MedCLIP+GPT2, CLIP+GPT2) $>$ \textbf{low-quality general-purpose VLM} (Phi-3.5-Vision). This ordering is observed across both NLP and clinical metrics, with MedGemma-4B achieving the highest ROUGE-1 (0.5750), ROUGE-L (0.5691), METEOR (0.6152), CIDEr (0.5818), word-level F1 (0.5610), and BI-RADS accuracy (0.5582). Although Qwen2.5-VL-7B shows slightly better BLEU-1 and ROUGE-2, its lower CIDEr and BI-RADS accuracy indicate weaker alignment with clinically relevant content.
%각 VLM의 성능 도출 결과 분석
This performance pattern can be explained by the interplay between domain alignment and representational quality, particularly in the context of DMID’s small size and clinical specificity. Medical-specific VLMs such as MedGemma-4B benefit from pretraining on radiology-style data and domain-specific terminology, which improves their ability to preserve fine-grained lesion descriptors (e.g., “spiculated mass”, “architectural distortion”) and maintain BI-RADS consistency under limited fine-tuning data. High-quality generalist models like Qwen2.5-VL-7B possess strong generic visual–linguistic alignment but lack inherent exposure to mammography-specific structures and language, leading to fluent but occasionally incomplete or clinically imprecise reports. Modular pipelines (CLIP+GPT2, MedCLIP+GPT2) rely on separate encoders and decoders, which may limit cross-modal contextual integration, especially for multi-view reasoning. Low-quality or compact generalist VLMs such as Phi-3.5-Vision, with limited pretraining scale and weaker vision–language alignment, fail to capture the detailed radiologic semantics required for accurate mammography reporting.
%정성적 평가에 대한 분석. 할루시네이션 & 일관성 측면에서의 분석
Qualitatively, MedGemma-4B produces coherent multi-view narratives with minimal hallucinations, and most deviations from the ground truth involve benign normal-structure mentions, which are clinically harmless or potentially informative for patients. In contrast, Qwen2.5-VL-7B, despite fluent text generation, often omits critical lesion details (e.g., margins, microcalcifications) or hallucinates unsupported findings. Phi-3.5-Vision frequently misclassifies BI-RADS categories and defaults to generic descriptors, while both CLIP+GPT2 and MedCLIP+GPT2 struggle with consistent lesion terminology, yielding outputs unsuitable for clinical draft usage.
%최종 요약
Overall, these results demonstrate that in small, clinically specialized datasets like DMID, domain-specialized pretraining yields the largest performance gains, followed by high-capacity generalist models, while modular or low-resource architectures lag significantly due to limited multimodal integration and weaker clinical grounding.

\subsection{Limitations and Future Works}
%DMID 데이터셋의 규모, 라벨 주관성, 보고서 표현의 다양성, 평가 한계로 인해 모델의 일반화와 임상적 신뢰성이 제한됨.
While our AMRG framework achieves strong gains in both linguistic fidelity and clinical accuracy, several limitations remain that reflect the intrinsic challenges of mammography report generation identified in our analysis. First, the DMID dataset is relatively small and imbalanced across BI-RADS categories, which, combined with the partially subjective nature of BI-RADS and breast density labeling, may limit generalization to rare findings and diverse populations. Second, narrative variability—where radiologists use heterogeneous terminology for the same lesion type—introduces noise that can destabilize training and complicate evaluation. Third, occasional model hallucinations and unsupported statements pose potential patient safety concerns, and our current evaluation pipeline relies on surface-level NLP metrics that do not fully capture clinical correctness or lesion–report consistency.
%향후에는 데이터 다양성과 품질을 높이고, 병변-리포트 정합성을 정량화할 수 있는 평가 프레임워크를 개발하며, 사실성 향상을 위한 프롬프트 및 디코딩 개선을 추진할 예정.
To address these limitations, future work will pursue several directions. We plan to construct an expanded, multi-institutional dataset with improved class balance and richer linguistic diversity, incorporating explicit quality control to reduce annotation subjectivity. We will also develop a mammography-specific evaluation framework that combines standard NLP metrics with lesion-level agreement analysis, adapting report–lesion mapping methods similar to CheXbert~\cite{smit2020chexbert} for the mammography domain. This will enable objective measurement of whether generated reports accurately describe annotated findings. Finally, we will explore fact-aware decoding and prompt refinement strategies—such as lesion-aware prompting inspired by PromptMRG~\cite{jin2024promptmrg}—to reduce hallucinations and improve factual alignment, thereby enhancing the clinical trustworthiness of automated mammography reporting.
\section{Conclusion}
%MedGemma-4B에 LoRA를 적용해 최초의 mammo 보고서 생성 벤치마크를 제안했으며, 대규모 범용 VLM 대비 우수한 성능을 달성. 제안 모델은 일관되고 임상적으로 타당한 리포트를 생성하며, 향후 mammo report generation 분야에 충분히 기여할 것으로 기대할 수 있다는 내용 작성
In this study, we propose a first benchmark for AMRG framework by fine-tuning the MedGemma-4B model using parameter-efficient LoRA adapters. Our approach achieves state-of-the-art performance compared to larger general-purpose VLMs, demonstrating the strength of domain-specialized pretraining combined with lightweight tuning. Qualitative analysis further confirms that our model generates coherent and clinically grounded reports with minimal hallucinations. We believe that our contributions will foster future research on radiology report generation in low-resource, high-stakes clinical domains.

\section*{Funding}
This work was supported by the National Cancer Center Grant(NCC-2311350-3).

\section*{Data availability}
The Digital Mammography Dataset for Breast Cancer Diagnosis Research (DMID) used in this study is available at Figshare: \url{https://doi.org/10.6084/m9.figshare.24522883.v2}.

% \bibliographystyle{elsarticle-num}
% \bibliography{reference}

\newpage
\section*{Appendix}

\renewcommand{\thetable}{A.\arabic{table}}
\setcounter{table}{0}

\subsection*{DMID Dataset}
\label{app:DMID_Dataset}

Table~\ref{tab:birads_split} presents the detailed distribution of BI-RADS codes across the train, validation, and test splits in the DMID dataset. As expected in a screening mammography context, BI-RADS 1 (negative) and BI-RADS 3 (probably benign) cases dominate all subsets. The training set includes a wider range of diagnostic categories, including a non-trivial number of high-suspicion cases such as BI-RADS 4a, 4b, 4c, and 5, which enables the model to observe diverse pathological patterns during learning. The validation and test sets, while smaller, retain meaningful representation of both benign and malignant classes, particularly BI-RADS 4a through 4c, allowing for balanced and clinically relevant evaluation. A small number of rare or ambiguous entries (e.g., “3 and 5”) are also included to reflect labeling uncertainty occasionally encountered in real-world radiology datasets.

\begin{table}[ht]
\centering
\caption{BI-RADS code distribution across dataset splits in DMID.}
\label{tab:birads_split}
\footnotesize
\begin{tabular}{cccc}  % 모든 열 c (center) 정렬
\toprule
\textbf{BI-RADS Code} & \textbf{Train} & \textbf{Validation} & \textbf{Test} \\
\midrule
0 & 1 & 0 & 0 \\
1 & 157 & 30 & 22 \\
2 & 24 & 1 & 5 \\
3 & 109 & 10 & 9 \\
3 and 5 & 1 & 0 & 0 \\
4 & 3 & 0 & 0 \\
4a & 31 & 1 & 5 \\
4b & 26 & 0 & 5 \\
4c & 39 & 7 & 6 \\
5 & 16 & 2 & 0 \\
\midrule
 Total& 407& 51&52\\
 \bottomrule
\end{tabular}
\end{table}
\end{document}